\documentclass{aa501}
\usepackage{graphicx}
\usepackage[english]{babel}

\begin{document}

\title{Discovery of strong CIV absorption in the highest redshift
 quasar\thanks{Based on observations made with
 the Italian Telescopio
Nazionale Galileo (TNG) operated on the island of La Palma by the Centro
Galileo Galilei of the CNAA (Consorzio Nazionale per l'Astronomia e
l'Astrofisica) at the Spanish Observatorio del
Roque de los Muchachos of the Instituto de Astrofisica de Canarias.}}

\author{R. Maiolino\inst{1} \and F. Mannucci\inst{2}
 \and C. Baffa\inst{1} \and S. Gennari\inst{1}
 \and E. Oliva\inst{1,3}}

\offprints{R. Maiolino}

\institute{Osservatorio Astrofisico di Arcetri, Largo E. Fermi 5, 50125
Firenze, Italy
 \and
CAISMI-CNR, Largo E. Fermi 5, 50125 Firenze, Italy
 \and
Centro Galileo Galilei \& Telescopio Nazionale Galileo,
P.O. Box 565, 38700, S. Cruz de La Palma, Spain
  }

  \date{Received ; accepted }

\abstract{
We report the near-IR
 detection of a prominent CIV absorption in the rest-frame UV
spectrum of the most distant known QSO, SDSS J104433.04$-$012502.2,
at $z=5.80$.
This QSO was recently observed with XMM-Newton and it was found to be
notably X-ray weak.
The equivalent width
of the CIV absorption feature ($\sim$10\AA ) strongly supports the
idea that the
X-ray faintness of this QSO is due to heavy absorption by gas with a column
density $N_H > 10^{24}$cm$^{-2}$. The shape of the CIV feature suggests
that this is a Broad Absorption Line QSO. Although absorbed
by a huge column of gas, the observed continuum in the 0.9--2.4$\mu$m range
($\sim$1300--3500 \AA \ rest frame) exactly matches the template of
unabsorbed QSOs without invoking any reddening ($\rm E_{B-V}<0.08$ mag),
indicating that dust in the absorbing gas is either absent or composed of
large grains.
\keywords{Infrared: galaxies -- quasars: general -- quasars: absorption lines}
}

\titlerunning{CIV absorption in the highest redshift quasar}
\authorrunning{Maiolino et al.}

\maketitle

\section{Introduction}

Recent studies have found a significant population of QSOs whose soft X-ray
emission is much weaker, with respect to the optical-UV emission, compared
to what observed in ``classical'' QSOs (Elvis \cite{elvis}, Laor et al.
\cite{laor}, Yuan et al. \cite{yuan}, Risaliti et al. \cite{risaliti}).
In this class of objects
the X-ray emission is more than 10--30 times fainter than expected from the
optically selected population of QSOs. Although some authors have
ascribed the X-ray faintness to an intrinsically different spectral
energy distribution (SED), more recently evidence was
found that absorption by gas along our line of sight might be
responsible for the observed properties of these objects. Indeed,
Brandt et al. (\cite{brandt00}) identified a correlation between the presence
of deep resonant UV absorption features (especially CIV) and X-ray
weakness. In particular, many of the X-ray weak QSOs result to be
Broad Absorption Line (BAL) QSOs (i.e. quasars with prominent absorption
features blueshifted with respect to the resonant UV lines and ascribed
 to gas outflowing with velocities from
from $\sim 5000$ km s$^{-1}$ up to more than 30000 km s$^{-1}$).
Vice versa, most BAL QSOs appear to
be X-ray weak AGNs. Hard X-ray studies have shown that the X-ray
faintness of these objects extends to the 2--10 keV band, and in some cases
evidence for a photoelectric cutoff due to absorbing gas
with $N_H \sim 10^{22}-10^{23}$cm$^{-2}$ was found, thus
supporting the absorption scenario (Gallagher et al. \cite{gallagher99}, 
\cite{gallagher01}).
Risaliti et al. (\cite{risaliti})
have found, among a grism selected sample of QSOs,
that X-ray weak QSOs tend to have redder colors, which is ascribed to
dust reddening, again supporting the absorption scenario.

\begin{figure*}[!ht]
\centering
\includegraphics[width=14truecm]{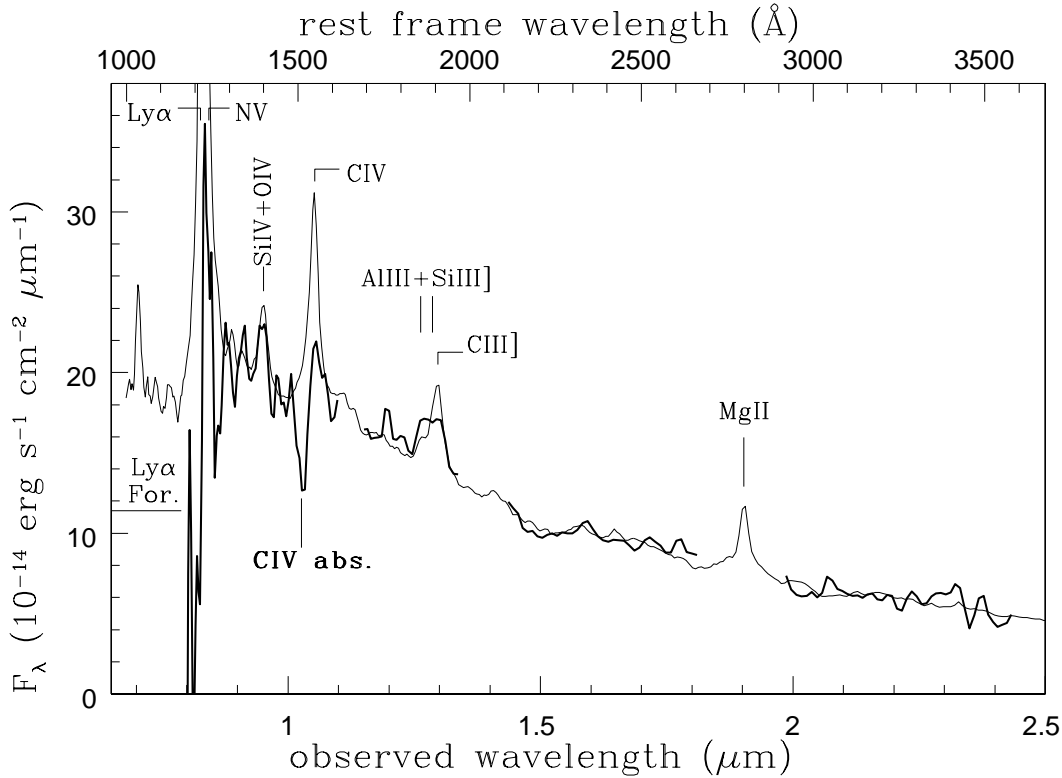}
 \caption{Thick line: observed spectrum of QSO SDSS 1044$-$0125.
 Thin line: template spectrum of (unabsorbed) optically selected
 QSOs (Francis et al. \cite{francis})
 smoothed to the same resolution of our spectrum.}
\end{figure*}

Most of the studies mentioned above (except for Risaliti et al.
\cite{risaliti})
deal with QSOs at moderate redshift ($z<1$). Recently, the
enhanced sensitivity of the X-ray XMM-Newton telescope
enabled to study the X-ray emission of the most distant QSOs.
In particular, Brandt et al. (\cite{brandt01a}) used XMM to
observe the recently discovered
QSO SDSSp J104433.04$-$012502.2 at $z=5.8$ (Fan et al. \cite{fan})
and achieved a detection
in the 0.5--2 keV band. When compared to the (rest frame)
UV emission, this high redshift QSO is X-ray weak with respect to the
optically selected QSOs, similarly to the lower redshift X-ray weak QSOs.
The nature of the X-ray weakness of this QSO is not clear and,
in analogy with the local X-ray weak QSOs, it could be ascribed either 
to an intrinsically different SED or to absorption
along our line of sight. As suggested by Brandt et al. (\cite{brandt01a})
near-IR observations aimed at detecting resonant absorption lines in
the rest-frame UV spectrum could tackle the issue.

In this letter we report new near-IR spectroscopic observations of this
high redshift QSO which strongly
support the scenario of heavy absorption due to gas along our line of sight.

\section{Observations}

The observations were obtained at the Italian Telescopio Nazionale Galileo,
a 3.56m telescope, with the Near Infrared Camera Spectrograph (NICS),
a cryogenic focal reducer designed as a near-infrared common-user instrument
for that telescope. The instrument is equipped with a Rockwell 1024$^2$ HAWAII
near infrared array detector. Among the many imaging and spectroscopic
observing modes (Baffa et al. \cite{baffa}),
NICS offers a unique, high sensitivity,
low resolution observing mode, which uses an Amici prims as a dispersing
element (Oliva \cite{oliva}).
In this mode it is possible to obtain the spectrum from 0.8$\mu$m
to $\sim$2.5$\mu$m in one shot. The spectral resolution with a $0.''75$ slit
(as it was in our case) is $\sim$75 and nearly constant over the whole
wavelength range. Clearly, this observing mode is an optimal tool
to study the near-infrared continuum of faint sources as well as
for the detection of broad ($\sim$5000 km s$^{-1}$) emission and absorption
lines in faint QSOs.

SDSSp J104433.04$-$012502.2 was observed on
December 9$^{th}$ and 12$^{th}$ 2000.
As mentioned above
we used a $0.''75$ slit, whose width projected on the array corresponds
to three pixels, yielding a spectral resolution of 4500 km s$^{-1}$.
Each of the two nights the object was
observed for 30 minutes. Unfortunately, the presence of
electronic noise during the second night prevented us to exploit the
data at wavelength shortward of $\sim$1.4$\mu$m of this second section
of observations, while at longer wavelengths we could combine the
data of both nights. Wavelength calibration was performed by
using an Argon lamp
and the deep telluric absorption features. The telluric absorption was then
removed by dividing the quasar spectrum by an A0 reference star spectrum
observed at similr airmass. The intrisic features and slope of the reference
star were then removed by multiplying the spectrum by the theoretical
spectrum of A0 stars smoothed to our resolution.

\section{Results and discussion}

In Fig.1 we plot the resulting spectrum of SDSS 1044$-$0125 (thick line).
The regions of bad atmospheric transmission are omitted.
The signal-to-noise is about 25 in the H band, 15 in the J and K
bands and 8 at 0.9$\mu$m.
The increased noise at $\lambda < 0.95\mu$m is due to the drop of the
detector sensitivity. However, both the
Ly$\alpha$ and the sharp break blueward of the Ly$\alpha$
are real (Fan et al. \cite{fan}); the latter is due
 to the Ly$\alpha$ forest.

The thin line is the QSO template obtained by Francis et al. (\cite{francis})
from a sample of optically selected QSOs,
normalized to match the continuum of SDSS 1044$-$0125 and smoothed to the
same velocity resolution of our spectrum.
It is quite impressive that the shape of SDSS 1044$-$0125 is essentially
identical to the (lower redshift) template. Also, the spectral slope
is $\alpha = -0.3$ (F$_{\nu}\propto \nu ^{\alpha}$)\footnote{This is the
slope in the range 1500--3500\AA , in analogy with the range adopted by
Francis et al. (\cite{francis}),
given that at shorter wavelengths various absorption systems affect the slope.}
which is consistent with 
the median slope found by Francis et al. for their sample of QSOs.
We will discuss these properties later on.

\begin{figure}[!ht]
\centering
\includegraphics[width=8truecm]{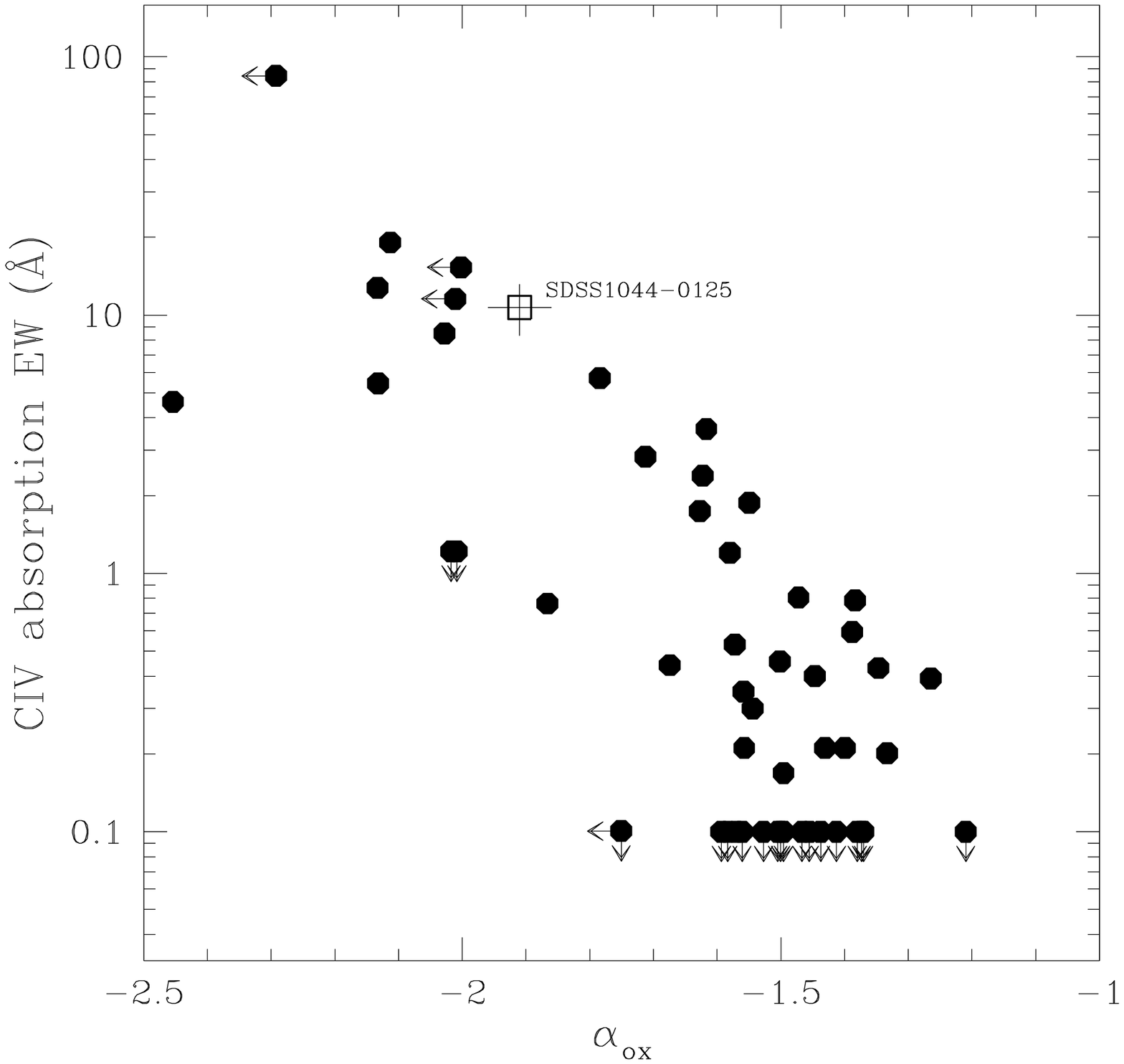}
 \caption{Equivalent width of the CIV absorption feature as
 a function of the optical--to--X-ray spectral index $\alpha_{ox}$ for
 a sample of QSOs (from Brandt et al. \cite{brandt00}). The location of
 SDSS 1044$-$0125 is identified with an hollow square.}
\end{figure}

Our spectrum shows evidence of the
emission of Ly$\alpha$+NV(1240\AA), SiIV+OIV($\sim$1400\AA ),
CIV (1549\AA ), and CIII] (1909\AA ) and possibly
AlIII (1857\AA ) and SiIII] (1892\AA ).
The MgII line at 2799\AA \ 
lies in the bad atmospheric transmission region between H and K.
Most interesting, we clearly
detect a CIV absorption feature blueshifted with
respect to the emission line, similarly to what observed in BAL QSOs
or in other X-ray weak QSOs at lower redshift.
The feature is marginally resolved: it has
a width of $\sim$7600 km s$^{-1}$ which implies an intrinsic width
(deconvolved from the instrumental resolution) of $\sim$6100 km s$^{-1}$.
The blue edge of the absorption feature extends to
$\sim$12000 km s$^{-1}$ from the CIV nominal wavelength.
These values are not uncommon among BAL QSOs. The rest-frame
equivalent width (EW) of the CIV absorption is
10.6$\pm$2.5\AA .

Regardless of whether SDSS 1044$-$0125 is actually a BAL QSO or not,
the detection of a deep CIV absorption feature strongly supports the
scenario that the X-ray weakness is to ascribe to heavy obscuration along
the line of sight. Indeed,
as discussed in the introduction, Brandt et al. (\cite{brandt00}) showed that 
CIV absorption is a powerful tool to identify QSOs whose X-ray faintness is
to ascribe to absorption by gas along the line of sight.
In Fig.2 we show a revised version of Fig.4 in Brandt et al. (\cite{brandt00}),
where for
a large sample of QSOs the EW of the CIV absorption line is reported
as a function of the optical--to--X-ray spectral index $\alpha _{ox}$
(i.e. the X-ray fainter sources have more negative values of $\alpha _{ox}$).
The location of SDSS 1044-0125 on this diagram is marked with a
hollow square and is in excellent agreement with the relation found
for the low redshift QSOs, thus supporting that the X-ray weakness of this
object is also due to gas absorption. As discussed in Brandt et al.
(\cite{brandt01a}),
within the absorption scenario
the lack of detection in the 2--7 keV band ($\sim$14--50 keV rest frame)
implies that the column of absorbing gas is larger than $N_H >
10^{24}$cm$^{-2}$, i.e. this QSO is Compton thick.

If the absorbing was characterized by a Galactic dust--to--gas ratio
and Galactic dust composition, then the optical extinction associated to the
X-ray absorption should be $\rm A_V > 500$~mag, which would completely
obscure the QSO in the optical and in the UV. Instead, not only the QSOs
is not optically absorbed, but it does not even appear to be reddened.
We estimate the maximum reddening to be $\rm E_{B-V} < 0.08$ mag
\footnote{The maximum reddening was estimated by assuming that the
intrinsic slope was as high as
$\alpha$=1 (only $\sim$10\% of the QSOs in the Francis et al.
(\cite{francis}) sample have a steeper slope, i.e. this is a deviation at about
2$\sigma$ from the median slope) and then determining the $E_{B-V}$
required to match the observed spectrum.}.
Such a mismatch between dust and gas absorption
was already noted previously in lower redshift objects (Maiolino et al.
\cite{maiolino01a}, Risaliti et al. \cite{risaliti}).
In this respect BAL QSOs (along with some
non-BAL QSOs) represent the extreme case of large gas absorption in the X-rays
and little or no dust extinction/reddening in the optical-UV. Within
this population of objects, SDSS 1044-0125 stands out as
an even more extreme case. It is
one of the BAL QSOs with the lowest reddening ever measured (Yamamoto \&
Vansevicius \cite{yamamoto}) and the XMM measurement, along with the high
redshift (which would shift a high energy
photoelectric cutoff into the soft band), allows
to set a lower limit on the absorbing column of gas which is the largest ever
set for this class of objects. These values constrain the $\rm A_V/N_H$
of the gas along the line of sight to be less than 10$^{-3}$ times the
Galactic standard value. The extremely reduced value of $\rm A_V/N_H$ 
might either indicate that dust in the absorbing gas is absent or that
it is mostly composed of large grains (which make the extinction curve
flatter and reduce $\rm A_V/N_H$, Maiolino et al. \cite{maiolino01b}).

Although the CIV absorption and the X-ray absorption appear related,
there is no clear indication that the two absorbing
gaseous media are physically associated. However, as noted by Brandt et al.
(\cite{brandt01b}),
if the X-ray absorbing gas has an $\rm N_H > 10^{24} cm^{-2}$ and
it is outflowing with the terminal velocities inferred by the CIV
absorption feature, this would imply a mass outflow larger than
$\rm 5~M_{\odot}yr^{-1}$ and an outflow kinetic energy larger than the
ionizing luminosity emitted by the QSOs. Should such large and massive
outflows result to be relatively common among high redshift QSOs, this
would have important implications on the chemical enrichment of the
intergalactic medium.

\section{Conclusions}

We have exploited a new low resolution near--infrared spectrometer
to obtain the spectrum of the most distant known QSO, SDSS 1044$-$0125,
over the whole 0.8--2.5$\mu$m wavelength range, corresponding to
$\sim$1170--3600\AA \ rest frame. The spectrum reveals a prominent CIV
blueshifted absorption feature, suggesting that
most likely this is a Broad Absorption Line QSO. The large equivalent
width of the CIV absorption feature (10.6\AA ) strongly supports that
the X-ray weakness, recently identified by means of XMM observations
of this QSO, is due to absorption by gas with $N_H > 10^{24}$cm$^{-2}$.

The continuum of this QSO exactly matches the template of unabsorbed,
optically selected QSOs, implying that no dust reddening is associated
to the gas responsible for the X-ray
absorption. We derive an upper limit to the
(Galactic-like) reddening of $\rm E_{B-V}<0.08$ mag,
which implies that
the $\rm A_V/N_H$ of the gas along the line of sight is less than
10$^{-3}$ of the Galactic value. The extremely reduced dust absorption and
reddening is either due to an extremely reduced dust content or to
dust mostly composed of large grains.

The inferred mass outflow is very high ($\rm 5~M_{\odot}yr^{-1}$).
If such large
outflows are common among high redshift QSOs, these
would have an important role in the chemical enrichment of the
intergalactic medium.

\begin{acknowledgements}
We are grateful to the Arcetri and TNG technical staff and to the TNG
operators for their assistance during the commissioning.
This work was partially supported by the
Italian Ministry for
University and Research (MURST) under grant Cofin00--02--36 and
by the Italian Space Agency (ASI) under grant 1/R/27/00.
\end{acknowledgements}


\begin{thebibliography}{}

 \bibitem[2000]{baffa}
Baffa, C., Gennari, S., Lisi, F., et al. 2000, in {\it The Scientific
 Dedication of the Telescopio Nazionale Galileo}, La Palma, November 3-5, 2000
 (astro-ph/0010328)

 \bibitem[2000]{brandt00}
Brandt, W.N., Laor, A., Wills, B.J., 2000, ApJ, 528, 637

 \bibitem[2001a]{brandt01a}
Brandt, W.N., Guainazzi, M., Kaspi, S., et al., 2001a, AJ, in press
 (astro-ph/0010328)

 \bibitem[2001b]{brandt01b}
Brandt, W.N., Gallagher, S.C., Laor, A., Wills, B.J., in
 X-ray Astronomy '999: Stellar Endpoints, AGNs
    and the Diffuse X-ray Background, in press (astro-ph/99103002)

 \bibitem[1992]{elvis}
Elvis, M. 1992, in Froentiers of X-ray Astronomy, ed. Tanaka, Y. \&
 Koyama, K. (Universal Acad. Press, Tokyo), p. 567

 \bibitem[2000]{fan}
Fan, X., White, R.L.,
  Davis, M., et al., 2000, AJ, 120, 1167

 \bibitem[1991]{francis}
Francis, P.J., Hewett, P.C., Foltz, C.B.,
  Chaffee, F.H., Weymann, R.J.,
  Morris, S.L., et al., 1991, ApJ, 373, 465

 \bibitem[1999]{gallagher99}
Gallagher, S.C., Brandt, W.N., Sambruna, R.M., Mathur, S.,
 Yamasaki, N., 1999, ApJ, 519, 549

 \bibitem[2001]{gallagher01}
Gallagher, S.C., Brandt, W.N., Laor, A., Elvis, M., Mathur, S.,
 Wills, B.J., Iyomoto, N., 2001, ApJ, 546, 795

\bibitem[1997]{laor}
 Laor, A., Fiore, F., Elvis, M., Wilkes, B.J.,
 McDowell, J.C., 1997, ApJ, 477, 93

 \bibitem[2001a]{maiolino01a}
Maiolino, R., Marconi, A., Salvati, M., et al., 2001, A\&A, 365, 28

 \bibitem[2001b]{maiolino01b}
Maiolino, R., Marconi, A., Oliva, E., 2001, A\&A, 365, 37

 \bibitem[2001]{oliva}
Oliva, E., 2001, Mem. Sc. Astr. It., in press (astro-ph/9909108)

 \bibitem[2001]{risaliti}
Risaliti, G., Marconi, A., Maiolino, R., Salvati, M., Severgnini, P.,
  2001, A\&A, in press (astro-ph/0102427)

 \bibitem[1999]{yamamoto}
Yamamoto, T.M., Vansevicius, V., 1999, PASJ, 51, 405

 \bibitem[1998]{yuan}
Yuan, W., Brinkmann, W.,
 Siebert, J., Voges, W., 1998, A\&A, 330, 108

\end{thebibliography}
\end{document}